\documentclass[%
 aip,onecolumn,
 amsmath,amssymb,
 reprint,%
]{revtex4-2}

\usepackage{graphicx}
\usepackage{dcolumn}
\usepackage{bm}

\usepackage[utf8]{inputenc}
\usepackage[T1]{fontenc}
\usepackage{mathptmx}
\usepackage{etoolbox}

\usepackage{color,graphicx,shortvrb,epsfig}

\usepackage{graphicx}
\usepackage{dcolumn}
\usepackage{bm}

\usepackage{colordvi}
\usepackage{epsfig,amssymb,amsmath,latexsym}

\def\lsim{\mathrel{\rlap{\lower4pt\hbox{\hskip1pt$\sim$}}
    \raise1pt\hbox{$<$}}}         
\def\gsim{\mathrel{\rlap{\lower4pt\hbox{\hskip1pt$\sim$}}
    \raise1pt\hbox{$>$}}}         

\def\lsim{\mathrel{\rlap{\lower4pt\hbox{\hskip1pt$\sim$}}
    \raise1pt\hbox{$<$}}}         
\def\gsim{\mathrel{\rlap{\lower4pt\hbox{\hskip1pt$\sim$}}
    \raise1pt\hbox{$>$}}}         

\def\beq{\begin{equation}}
\def\eeq{\end{equation}}
\def\ba{\begin{eqnarray}}
\def\ea{\end{eqnarray}}

\def\<{\langle}
\def\>{\rangle}

\begin{document}
\preprint{AIP/123-QED}

\begin{flushright}
\date{}
\end{flushright}

\preprint{AIP/123-QED}

\title{Chain-Length-Dependent Partitioning of 1-Alkanols in Raft-Like Lipid Membranes}
\author{Anirban Polley} 
\newcommand{\affA}{Raman Research Institute, C.V. Raman Avenue, Bangalore 560080, India}
\newcommand{\affB}{National Centre for Biological Sciences (TIFR), Bellary Road, Bangalore 560065, India}
\newcommand{\affC}{Tampere University of Technology, Korkeakoulunkatu 10, 33720 Tampere, Finland}
\newcommand{\affD}{SASTRA University, Tirumalaisamudram, Thanjavur, Tamilnadu-613401, India}

\affiliation{\affD}
 \email{anirban.polley@gmail.com}


\date{\today}




\begin{abstract}

Although 1-alkanols are widely used as anesthetics and membrane-active agents, the molecular basis of their chain-length-dependent cutoff behavior remains unclear. Here, we perform extensive atomistic molecular dynamics simulations to investigate the partitioning of 1-alkanols with varying chain lengths in a raft-like lipid bilayer composed of dipalmitoylphosphatidylcholine (DPPC), dioleoylphosphatidylcholine (DOPC), and cholesterol (Chol), which exhibits coexistence of liquid-ordered ($l_o$) and liquid-disordered ($l_d$) domains. We observe pronounced lateral heterogeneity in alkanol distribution, membrane thickness, number density, and lateral pressure profiles across coexisting phases. A distinct cutoff chain length, $n_{cutoff}=12$, is identified: alkanols with $n<n_{cutoff}$ preferentially partition into DOPC-rich $l_d$ domains, whereas alkanols with $n \ge n_{cutoff}$ preferentially localize within DPPC- and cholesterol-rich $l_o$ domains. This chain-length-dependent redistribution is accompanied by systematic reductions in the lateral pressure profile, membrane compressibility, and bending rigidity of the bilayer. The results provide a detailed molecular characterization of how alkanol chain length modulates membrane structure and mechanical response in laterally heterogeneous lipid membranes.

\end{abstract}

\maketitle

\section{\label{sec:level1}Introduction}
\noindent
The phenomenon of general anesthesia has been known for a long time \cite{seeman_anesthetic}, and the use of anesthetics in hospitals is routine for painless surgical procedures. However, the molecular-level mechanism underlying general anesthesia remains poorly understood. Anesthetics are known to alter membrane properties such as the total membrane volume, the volume occupied by anesthetic molecules within the membrane, the phase transition temperature, lipid chain order, membrane thickness, and the lateral pressure profile of the membrane \cite{Regen_2009,Ueda_1998,Ueda_2001}.
Despite extensive investigation, the mechanism of anesthesia remains enigmatic, particularly regarding how general anesthetics interact with specific proteins \cite{Frank_1994,Frank_1997} and induce conformational changes that lead to anesthetic effects.

In addition to their anesthetic effects, alkanols are widely used as penetration enhancers in transdermal drug delivery. However, the molecular mechanism governing their penetration-enhancing action remains unclear. Notably, 1-alkanols exhibit a pronounced cutoff effect when acting either as anesthetics or as penetration enhancers. The anesthetic potency of 1-alkanols increases with increasing chain length up to a critical cutoff at dodecanol, beyond which anesthetic activity is lost \cite{miller_anesthetic}. A similar trend is observed for penetration enhancement, where efficacy increases with chain length up to decanol and decreases for longer-chain 1-alkanols. Furthermore, branching of the carbon chain reduces the potency of 1-alkanols in both anesthetic and penetration-enhancing roles. 

A range of experimental and computational studies has examined the effects of anesthetics on lipid membrane properties. NMR spectroscopic measurements indicate that ethanol induces disorder in lipid acyl chains \cite{Feller_2002,Holte_1997,Barry_1995,Patra_2006,Joaquim_2011,Igor_2012}, while X-ray scattering studies suggest that ethanol can significantly modify membrane structure, particularly above the main phase transition temperature \cite{Cantor_1997a,Cantor_1997b,Cantor_1998,suryabrahmam2025,Ingolfsson2011}. Complementary molecular dynamics simulations have further provided molecular-level insight into anesthetic–membrane interactions and their impact on bilayer structure and dynamics \cite{Patra_2006,Bandyopadhyay_2004,Bandyopadhyay_2006,Smit_2004,Terama_2008,Griepernau_2007,Faller_2007,Vierl_1994,Jackson_2007,Gawrisch_1995,Dunn_1998,Mcintosh_1984}.

Despite extensive experimental and computational efforts, a unified molecular-level understanding of anesthetic and penetration-enhancing actions—particularly the origin of the cutoff effect—remains elusive. Anesthetic-induced changes in membrane structure and dynamics are expected to modify the lateral pressure profile, area compressibility modulus, and bending rigidity, as well as the associated free-energy landscape of the lipid bilayer, thereby influencing membrane-mediated protein function. In this context, 1-alkanols constitute an ideal model system, as their anesthetic and penetration-enhancing potencies exhibit a pronounced dependence on chain length and molecular architecture. Here, we investigate a series of 1-alkanols spanning a broad range of chain lengths in a raft-like model membrane composed of DOPC, DPPC, and cholesterol, which exhibits phase coexistence of liquid-ordered ($l_o$) and liquid-disordered ($l_d$) domains. This model system enables us to examine how alkanol incorporation alters membrane structural heterogeneity, cluster formation, penetration and partitioning behavior, and mechanical properties inferred from pressure-tensor and area-fluctuation analyses \cite{simons,simons_science2010,simons_nat_rev_2000,anirban_cell15,TrafficRaoMayor,raftreviews,sharma,debanjan,kripa_2012,anirban_pre18}. By systematically correlating these membrane perturbations with alkanol chain length, this study aims to provide mechanistic insight into the molecular origin of the cutoff effect and its implications for anesthetic and penetration-enhancing activity.

The remainder of the article is organized as follows. We first describe the atomistic molecular dynamics simulations of the multicomponent lipid bilayer. We then present our main results, focusing on spatial heterogeneity, cluster formation, alkanol penetration and partitioning, lateral pressure profiles obtained from the pressure tensor, area compressibility moduli derived from area fluctuations, bending rigidity inferred from mechanical stress distributions, and lipid order parameters. Finally, we summarize the key findings and discuss their broader implications.

\section{\label{sec:level2}Methods}
\noindent
{\it Model membrane}\,:\,

We study a symmetric three-component lipid bilayer membrane embedded in an aqueous environment using atomistic molecular dynamics (MD) simulations performed with GROMACS 5.1 \cite{Lindahl}. The bilayer is prepared at a temperature of $23^{\circ}C$ with relative molar concentrations of $33.3\%$ DOPC, $33.3\%$ DPPC, and $33.3\%$ cholesterol. To this symmetric ternary bilayer, 1-alkanols with varying chain lengths, $n=2$, $5$, $8$, $10$, $12$ and $16$, are introduced into the aqueous phase on both sides of the membrane at a concentration corresponding to $25\%$ relative to the lipid content.

Each multicomponent bilayer system consists of $512$ lipids per leaflet ($1024$ lipids in total) and $32768$ water molecules, corresponding to a water-to-lipid ratio of $32:1$, ensuring full hydration of the membrane. The chosen membrane composition lies within the phase-coexistence region and exhibits liquid-ordered ($l_o$) and liquid-disordered ($l_d$) domains \cite{schwille_dopc_dppc_chol}. \\

\noindent
{\it Force fields}\,:\,

Force-field parameters for DOPC, DPPC, and cholesterol are taken from previously validated united-atom lipid models \cite{kindt_dopc_dppc_chol,anirban_pre2025,anirban_jpcb12,anirban_cpl13,Tieleman-POPC,mikko}. Parameters for 1-alkanols with varying chain length are adopted from established and widely used models \cite{bockmann_alkanol,anirban_cpl13,MikkoBPJ2006,ramon_jpcb2011,ramon_plosone2013,anirban_pre2025}. Water molecules are modeled using the SPC/E water model, which includes an average polarization correction to the potential energy.    
\\

\noindent
{\it Initial configurations}\,:\,

Initial configurations of the symmetric multicomponent bilayer membranes are generated using PACKMOL \cite{packmol}, with all lipid components homogeneously mixed. Each bilayer contains $342$ DOPC, $340$ DPPC, and $342$ cholesterol molecules, randomly and uniformly distributed in space, and is hydrated with $32768$ water molecules. A total of $256$ 1-alkanol molecules are placed symmetrically in the aqueous regions on both sides of the bilayer using PACKMOL. A representative initial configuration is shown in Fig. S1 of the Supplementary Information (SI).\\

\noindent
{\it Simulation protocol and equilibration}\,:\,

Each system is first equilibrated for $50$\,ps in the NVT ensemble using a Langevin thermostat to remove unfavorable steric contacts. This is followed by production simulations of $2000$\,ns in the NPT ensemble at $T = 296$\,K ($23^{\circ}$C) and $P =1$\,atm. To improve statistical reliability, four independent simulations are performed for each system, yielding a total simulation time of $8 \mu s$ per membrane composition.

During the first $100$\, ns of NPT simulations, the Berendsen thermostat and barostat are used for further equilibration. Subsequently, the Nosé–Hoover thermostat and Parrinello–Rahman barostat are employed to generate the correct ensemble, with semi-isotropic pressure coupling and a compressibility of $4.5\times 10^{-5}$ bar$^{-1}$. Long-range electrostatic interactions are treated using the reaction-field method with a cutoff of $r_c = 2$\,nm, while Lennard–Jones interactions are truncated at $1$\,nm \cite{anirban_jpcb12,mikko,patra2004}. The final $500$\,ns of each trajectory are used for analysis.\\

\noindent
{\it Computation of rigidity}\,:\, 

The lateral pressure profiles across the bilayer are computed using the Irving–Kirkwood contour with a spatial grid resolution of $0.1$\,nm. Pairwise force contributions required for the pressure tensor are obtained by rerunning the trajectories with a cutoff of $2$\,nm for electrostatic interactions. Bond lengths are constrained using the LINCS algorithm for lipids and the SETTLE algorithm for water molecules, allowing the use of an integration time step of $2$\,fs \cite{Hess,SETTLE}.

Pressure profiles are accumulated over the last $500$\,ns of the trajectories. The lateral stress profile is then used to compute the bending rigidity of the membrane via the second moment of the pressure profile with respect to the bilayer midplane \cite{samuli_prl2009}. This approach provides a mechanical route to bending rigidity that directly links local stress distributions to membrane elastic properties. 


\section{\label{sec:level3}Results and Discussion}

To verify that the bilayer membranes are thermally and chemically equilibrated, we examine the time evolution of the total energy and the area per lipid (Fig. S2 in the SI). Both quantities exhibit asymptotic behavior with only small fluctuations, indicating that the systems have reached equilibrium \cite{anirban_jcb14,anirban_jpcb12,anirban_cpl13}.

\subsection{Mechanical properties of the membrane}

Mechanical equilibration and stability of the membranes are assessed through analysis of the stress profiles. For this purpose, the entire production trajectories from four independent simulations of each system are rerun and divided into approximately $\sim 125\times125\times72$ spatial grids with a grid spacing of $0.1$\,nm.

The pressure of a system can be written from the virial by the tensor $P^{\alpha \beta}(r)$, ($\alpha$, $\beta \in \{{x,y,z}\}$) as function of the position {\bf r}. It can be written as ,

\begin{equation}
P^{\alpha \beta}(r)=\sigma^{\alpha \beta}_{K}(r)+\sigma^{\alpha \beta}_{C}(r)
\label{tensorP}
\end{equation}

Here, $\sigma_{K}(r)$, the kinetic part of the stress tensor denotes the flux of momentum, defined as,

\begin{equation}
\sigma_{K}^{\alpha \beta}(r,t)=\langle \sum_{i=1}^{N} m_{i}v_{i}^{\alpha}v_{i}^{\beta}\delta(r-r_{i}) \rangle
\label{tensorK}
\end{equation}

where, the angular bracket $\langle ... \rangle$ refers the ensemble average, $N$ is the the total number of the particles in the system, 
 $m_i$, $r_i$ and $v_i$ are mass, position and velocity of the $i^{th}$ particle in the system.
 
The other component of the Eqn. \ref{tensorP}, the configurational part of the stress tensor describes the pairwise interactions which can be written as, 

\begin{equation}
\sigma_{C}^{\alpha \beta}(r,t)=\frac{1}{2} \langle \sum_{i}\sum_{j<i}F_{ij} \oint_{C_{ij}}dl^{\beta}\delta(r-l) \rangle 
\label{tensorC}
\end{equation}

where, the prefactor $\frac{1}{2}$ accounts for the double counting of the pairwise interaction; $C_{ij}$ takes account of contour between $i^{th}$ particle to $j^{th}$ particle and $F_{ij}$ is force acting on $i^{th}$ particle by $j^{th}$ particle. From the Eqn. \ref{tensorP}, \ref{tensorK} and \ref{tensorC}, we get the local pressure $P_v$ at any small volume V by taking average as $P_v=\int_{v} dr \frac{P(r)}{V}$ where Irving-Kirkwood contour has been used. 

The lateral pressure profile of the bilayer membrane is defined as, 

\begin{equation}
\pi(z)=P_{L}(z)-P_{zz}(z)
\label{lat_press}
\end{equation}

where, 
$P_{L}=\frac{P_{xx}(z)+P_{yy}(z)}{2}$, is lateral (in-plane) stress components and $P_{zz}$ is normal (out-of-plane) stress component averaged over the membrane plane. The coordinate z is the normal to the membrane.

Assuming the planner symmetry of the membrane, the surface tension of the bilayer membrane between $z_1$  and $z_2$ is defined as,

\begin{equation}
\gamma=\int^{z_{2}}_{z_{1}}dz \pi(z)
\label{gamma}
\end{equation}

According to Helfrich, the surface energy due to the curvature per unit area can be written as,

\begin{equation}
g(C_{1},C_{2})=\frac{1}{2}\kappa (C_1+C_2-C_0)^2+\kappa_{G}C_1 C_2
\label{FE}
\end{equation}

where,
 $C_1$ and $C_2$ are the local principal curvatures and $C_0$ is the spontaneous curvature, $\kappa$ and $\kappa_{G}$ are the bending rigidity and elastic modulus for Gaussian curvature respectively. The $\kappa C_0$ and $\kappa_{G}$ can be calculated from the $1^{st}$ and $2^{nd}$ moments of the lateral pressure profile as in Eqn. \ref{KC} and \ref{KG} respectively.

\begin{equation}
\kappa C_0=\int_{0}^{h} dz(z-\delta)\pi(z)
\label{KC}
\end{equation}

\begin{equation}
\kappa_{G} = \int_{0}^{h} dz(z-\delta)^2 \pi(z)
\label{KG}
\end{equation}

Moreover, the net force and net torque acting on the membrane are defined, respectively, as $F^\alpha=\int \partial_k P^{\alpha k} dv$ and $M^{\alpha \beta}=\int (\partial_l P^{\alpha l} x_\beta-\partial_l P^{\beta l} x_\alpha )dv$. 
Here, $h$ is the thickness of the monolayer where, $z=0$ is the center of the bilayer and $z$ is the normal coordinate of the bilayer. $\delta$ is the coordinate of the neutral plane of the monolayer.

For all eight systems considered—namely, the ternary bilayer without 1-alkanols and with 1-alkanols of chain lengths ($n=2$, $5$, $8$, $10$, $12$, $14$ and $16$)—the mean surface tension $\gamma$,  calculated over the last $500$\,ns of the trajectories, is found to be zero within statistical uncertainty (Fig. S3 in the SI), confirming tensionless bilayer conditions. Furthermore, the mean values of the net force components ($F_x$, $F_y$, $F_z$) and net torque components ($M_{xy}$, $M_{yz}$, $M_{zx}$) are also zero for all systems (Fig. S4 in the SI), demonstrating that the membranes are mechanically stable and fully equilibrated.

\begin{figure*}[h!t]
\begin{center}
\includegraphics[width=18.0cm]{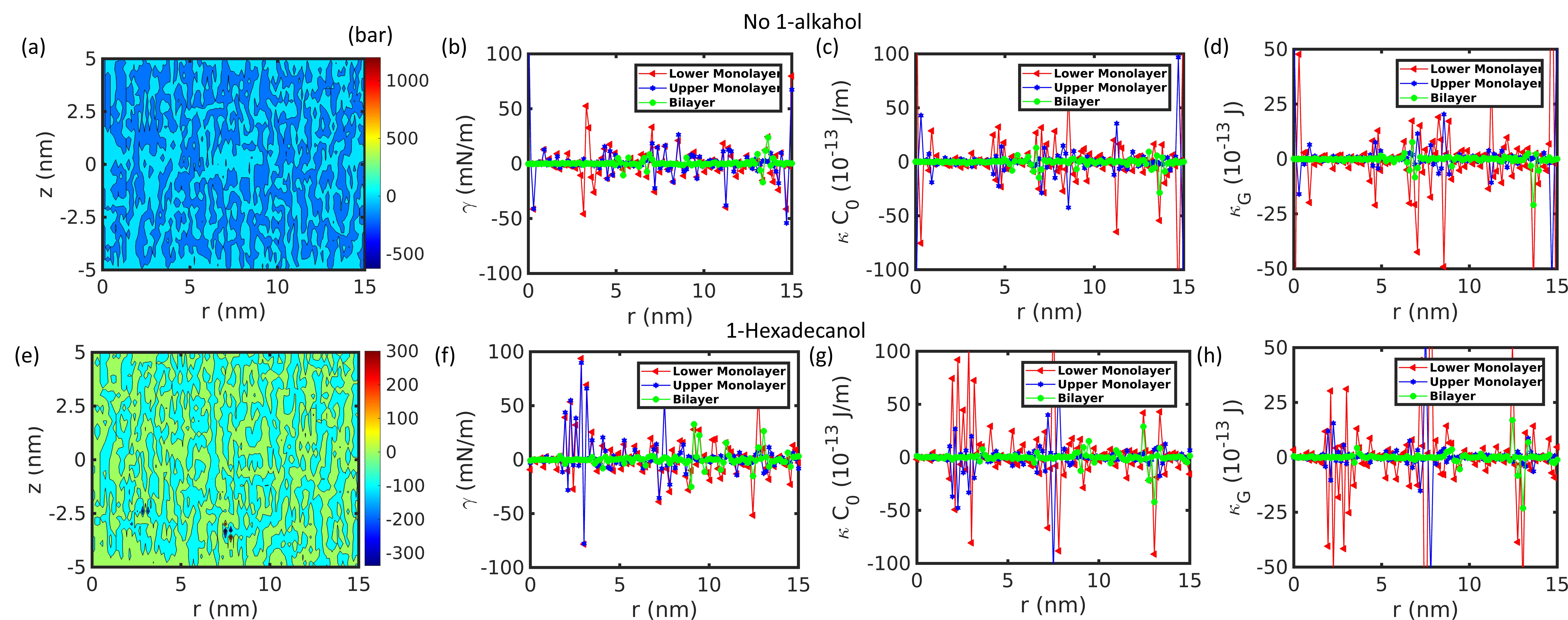}
\caption{Lateral pressure profile and its moments for the composite membrane in the absence (upper panels) and presence (lower panels) of 1-hexadecanol. The pressure profile, surface tension ($\gamma$), first moment (product of bending rigidity $\kappa$ and spontaneous curvature $C_0$), and second moment (Gaussian curvature modulus $\kappa_G$) are calculated using a cylindrical coordinate representation, with the z-axis normal to the bilayer plane. Panels (a) and (e) show the spatial heterogeneity of the lateral pressure profile for the composite membrane without and with 1-hexadecanol, respectively, illustrating a reduction in pressure upon alkanol incorporation. The radial dependence of $\gamma$ is shown in panels (b) and (f), that of  $\kappa C_0$ in panels (c) and (g), and that of $\kappa_G$ in panels (d) and (h), for the upper leaflet, lower leaflet, and the full bilayer. The corresponding radial averages of  $\kappa C_0$ and $\kappa_G$ are summarized in Table-I, indicating a reduction in membrane rigidity in the presence of alkanols. Mechanistically, alkanol incorporation redistributes lateral stresses across the bilayer, leading to membrane softening through reduced elastic moduli.
}
\label{fig1}
\end{center}
\end{figure*}

We employ a cylindrical geometry \cite{shi_jcp2023} to compute the pressure profile of the membrane and to examine its variation along the membrane normal ($z$-axis) as well as as a function of the radial distance ($r$) from the bilayer center (see Fig. S3 of the SI). Within this framework, we evaluate the surface tension ($\gamma$) and the first and second moments of the lateral pressure profile. The first moment corresponds to the product of the bending rigidity and the spontaneous curvature ($\kappa C_0$), while the second moment is related to the Gaussian curvature elastic modulus ($\kappa_{G}$). Analysis as a function of the radial coordinate, $r$ reveals tensionless membranes and a systematic reduction in the lateral pressure profile, $\kappa C_0$, and $\kappa_{G}$ in the presence of 1-alkanols compared to the alkanol-free membrane, as shown in Fig. \ref{fig1} and Table-I.

\begin{table*}[h!t]  
\begin{center}
\begin{tabular}{|c|c|c|}   \hline
System                    &   $\kappa C_o$  ($10^{-12} J/m$)     &  $\kappa_{G}$ ($10^{-12} J$) \\  \hline
membrane without alkanol           &            $5.012\pm 12.758$   &    $2.3714\pm 2.9067$  \\  \hline
membrane with 1-Ethanol              &         $3.2609\pm7.878$      &      $1.0845\pm3.7158$     \\  \hline
membrane with 1-Pentanol             &       $3.0247\pm 4.8748$        &       $0.652\pm6.2408$   \\  \hline
membrane with 1-Octanol              &      $1.8989\pm6.2768$         &     $0.50771\pm13.409$ \\  \hline
membrane with 1-Decanol              &      $0.50265\pm10.835$         &    $0.39726\pm3.2174$ \\  \hline
membrane with 1-Dodecanol            &       $0.30923\pm4.8056$        &     $0.1274\pm2.4468$   \\  \hline
membrane with 1-Tetradecanol         &      $-0.038644\pm5.2737$         &   $0.10918\pm3.185$\\  \hline
membrane with 1-Hexadecanol          &      $0.0011\pm11.085$         &  $-0.02814\pm2.6915$\\  \hline
\end{tabular} 
\label{tab1}
	\caption{First moment ($\kappa C_0$, product of bending rigidity $\kappa$ and spontaneous curvature $C_0$) and second moment ($\kappa_G$, Gaussian curvature modulus) of the lateral pressure profile for the composite bilayer membrane in the absence and presence of 1-alkanols. The results reveal a reduction in both $\kappa C_0$ and $\kappa_G$ upon alkanol incorporation, indicating that 1-alkanols soften the membrane and lower its resistance to bending and Gaussian deformations, consistent with their chain-length–dependent partitioning and pressure-profile modulation.
}
\end{center}
\end{table*}

\subsection{Transverse heterogeneity of the components in membrane}

\begin{figure*}[h!t]
\begin{center}
\includegraphics[width=16.0cm]{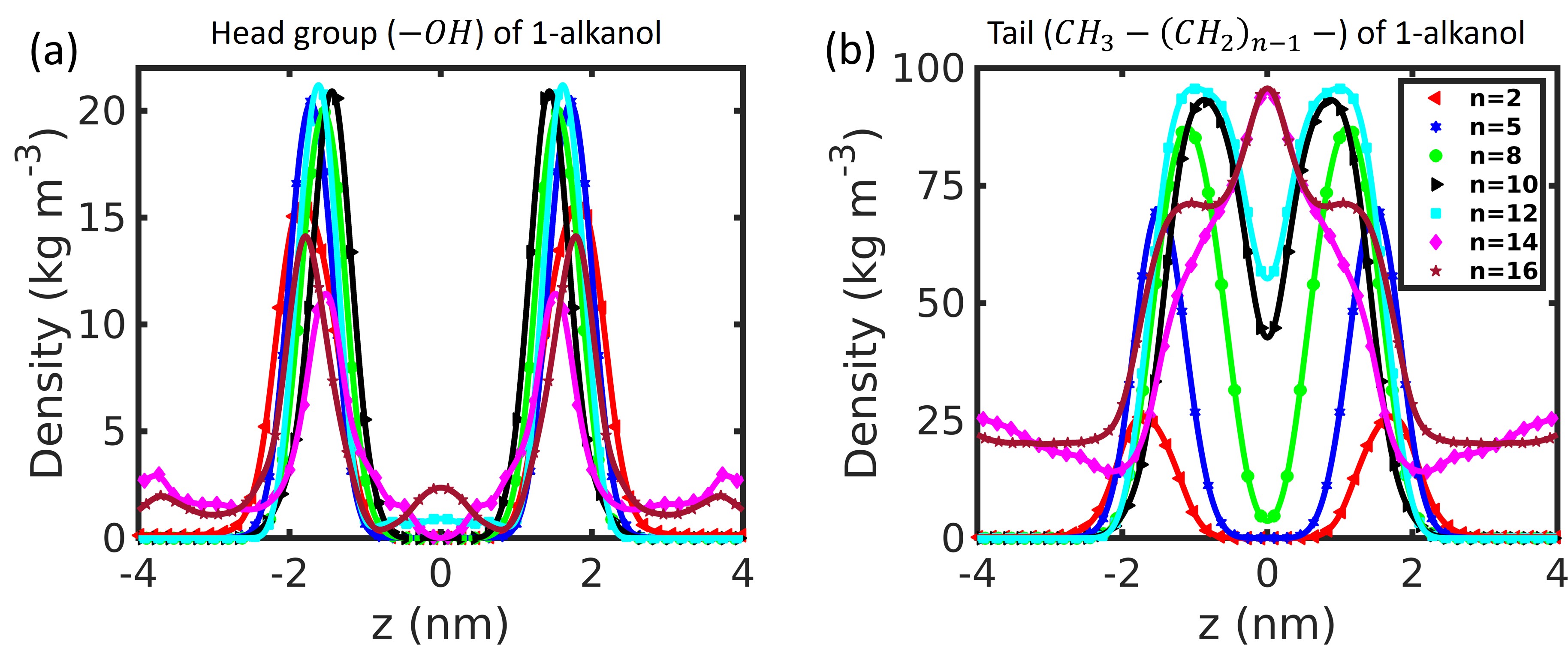}
\caption{Density profiles of the headgroup and tail of 1-alkanols in the composite bilayer membrane composed of DOPC, DPPC, and cholesterol. The z-coordinate distributions of the 1-alkanol headgroup (–OH) and tail (acyl chain, $-(CH2)_{n-1} -CH_3$, with chain length $n$) are computed from the last $500$\,ns of four independent $2000$\,ns ($2\,\mu s$) simulations for each system. Panels (a) and (b) show the density profiles of the headgroups and tails, respectively, from which it is evident that 1-alkanols penetrate progressively deeper into the membrane with increasing chain length. Therefore, deeper insertion of longer-chain alkanols enhances coupling to the bilayer core, leading to redistribution of lateral stresses and a reduction in membrane elastic moduli.
}
\label{fig2}
\end{center}
\end{figure*}

The density profiles of 1-alkanols are shown in Fig. \ref{fig2}. Specifically, the densities of the headgroup (($-OH$) and the hydrocarbon tail ($CH_3-(CH_2)_{n-1}-$) of the 1-alkanols are plotted as functions of the z-coordinate, where the z-axis is defined as the membrane normal and the bilayer lies in the xy-plane.

Fig. \ref{fig2} (a) shows two pronounced peaks in the headgroup ($-OH$) density profile, symmetrically located at $z \approx \pm 2.058$ \,nm. These peaks indicate that the hydroxyl groups of the 1-alkanols preferentially reside near the lipid head–neck region of the bilayer. In contrast, Fig. \ref{fig2} (b) displays the density profiles of the hydrocarbon tails, from which the penetration depth of the 1-alkanols into the membrane can be inferred. Here, the penetration depth is defined as the distance between the phosphate plane of the lipid headgroups and the equilibrated center-of-mass position of the alkanol tail. The results indicate that the penetration depth increases systematically with increasing alkanol chain length, $n$. 

In addition, the density profiles of the head and tail groups of each membrane component (DPPC, DOPC, and cholesterol), together with those of the 1-alkanols, are shown in Fig. S5 of the SI. From these profiles, we determine the membrane thickness, defined as the distance between the average positions of the phosphate planes of the two opposing leaflets. The calculated membrane thickness shown in Fig. S4 in the SI reveal a systematic decrease in membrane thickness with increasing alkanol chain length $n$.

\subsection{Clustering of 1-alkanol in membrane}

\begin{figure*}[h!t]
\begin{center}
\includegraphics[width=14.0cm]{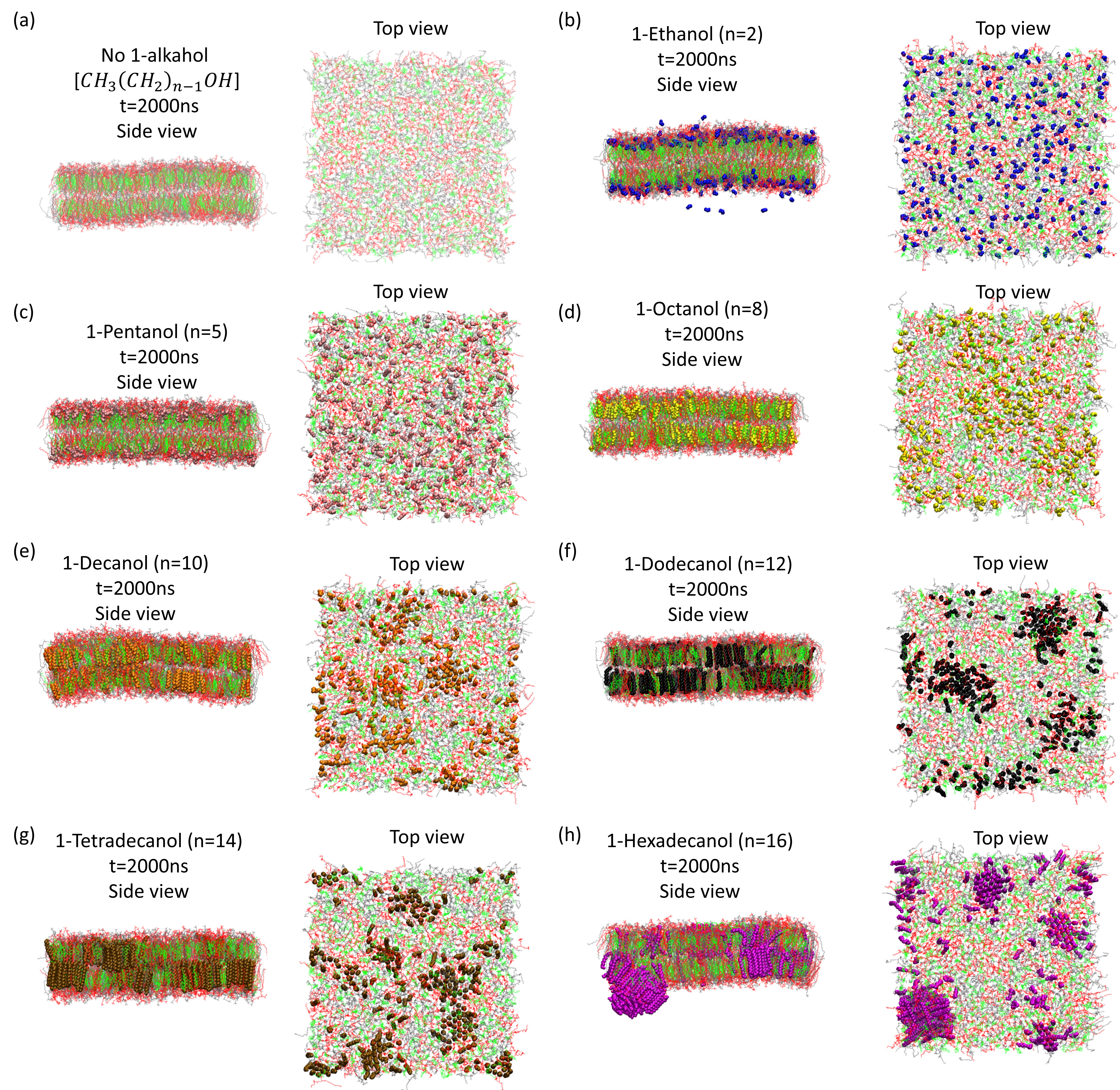}
\caption{Representative side and top views of the final configurations ($t=2000$\,ns) of symmetric composite bilayer membranes composed of DOPC (gray), DPPC (red), cholesterol (green), and water (cyan): (a) membrane without 1-alkanol; membranes containing (b) 1-Ethanol (blue), (c) 1-Pentanol (pink), (d) 1-Octanol (yellow), (e) 1-Decanol (orange), (f) 1-Dodecanol (black), (g) 1-Tetradecanol (dark orange), and (h) 1-Hexadecanol (purple). Thus, side views reveal progressively deeper membrane penetration with increasing alkanol chain length, while top views show domain-selective clustering of short-chain alkanols in DOPC-rich $l_d$ regions and long-chain alkanols in DPPC-rich $l_o$ regions.
}
\label{fig3}
\end{center}
\end{figure*}

The side/ top view of snapshot at time $t=2000$\, ns of the trajectory of composite bilayer membrane comprising DOPC, DPPC, Chol with and without 1-alkanols with varying chain length, $n$, respectively are shown in Fig. \ref{fig3} suggesting inhomogeneous distribution and clustering of the 1-alkanols in the membrane.

\begin{figure*}[h!t]
\includegraphics[width=10.0cm]{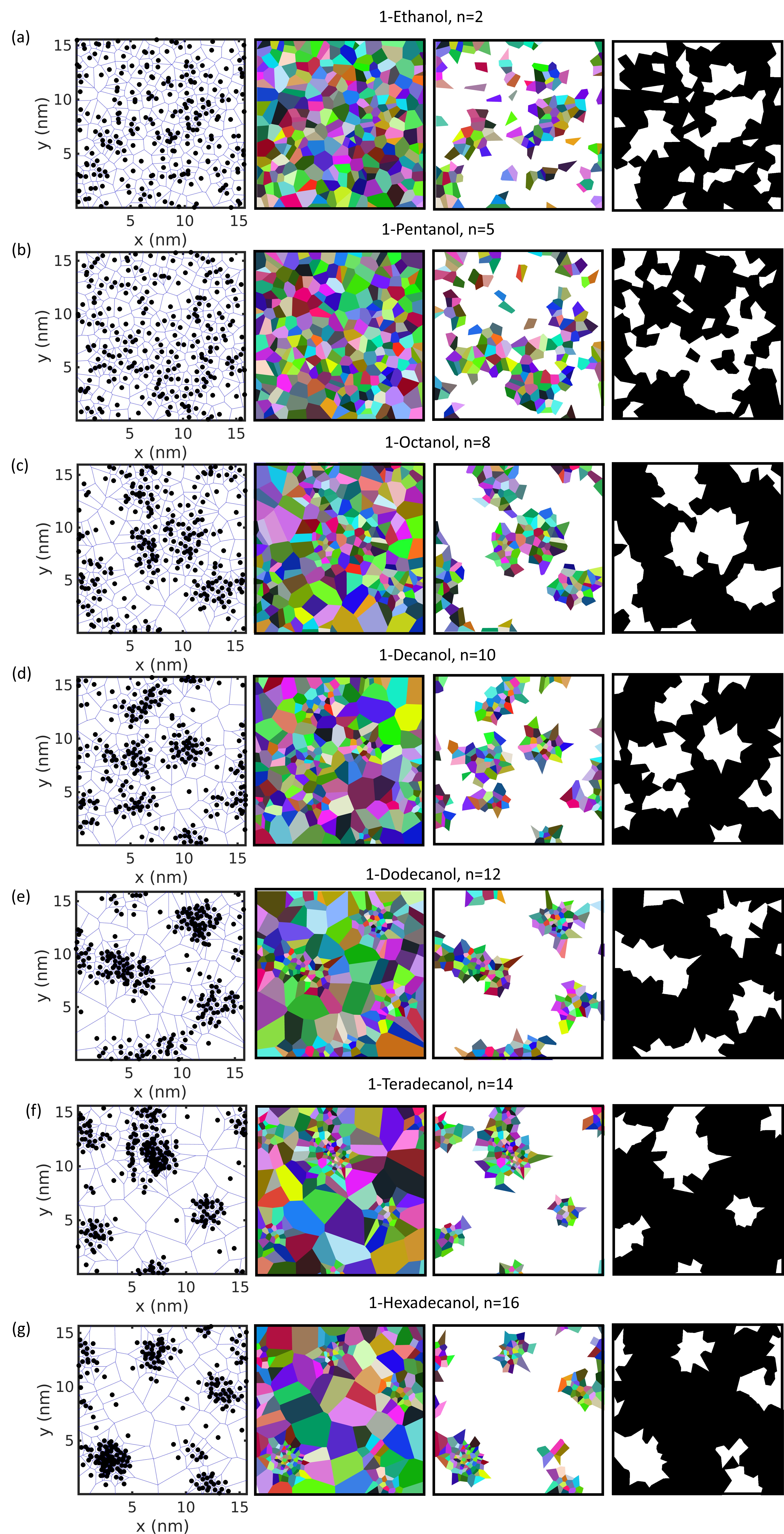}
\caption{Clustering of 1-alkanols in the composite bilayer membrane, identified using the Voronoi tessellation technique for varying acyl chain lengths. Black circles denote individual 1-alkanol molecules. Cluster analysis is performed using the last $500$\,ns of each trajectory ($5000$ frames) from four independent $2000$\,ns simulations per system. For each frame, the 
(x,y) coordinates of the center of mass (COM) of all 1-alkanol molecules are extracted. The analysis proceeds as follows: (i) spatial distributions of the COMs at a representative time ($t=2000$\,ns) are constructed using Voronoi tessellation (first column); (ii) Voronoi diagrams are generated, with distinct colors indicating individual Voronoi cells (second column); (iii) local number density is computed as the inverse of each Voronoi cell area, and cells with densities exceeding the mean density are selected (third column); and (iv) high-density cells are rendered in white, while low-density regions are shaded in black, yielding binary cluster maps (fourth column). High-resolution ($\sim 1024 p$ ) binary images are analyzed in MATLAB using the bwlabel function to quantify the area of polygonal white regions corresponding to individual 1-alkanol clusters. Briefly, the emergence of larger, more compact clusters with increasing chain length reflects enhanced membrane insertion and lateral pressure reduction, providing a membrane-mediated basis for the anesthetic cutoff phenomenon.
}
\label{fig4}
\end{figure*}

\begin{figure*}[h!t]
\begin{center}
\includegraphics[width=8.0cm]{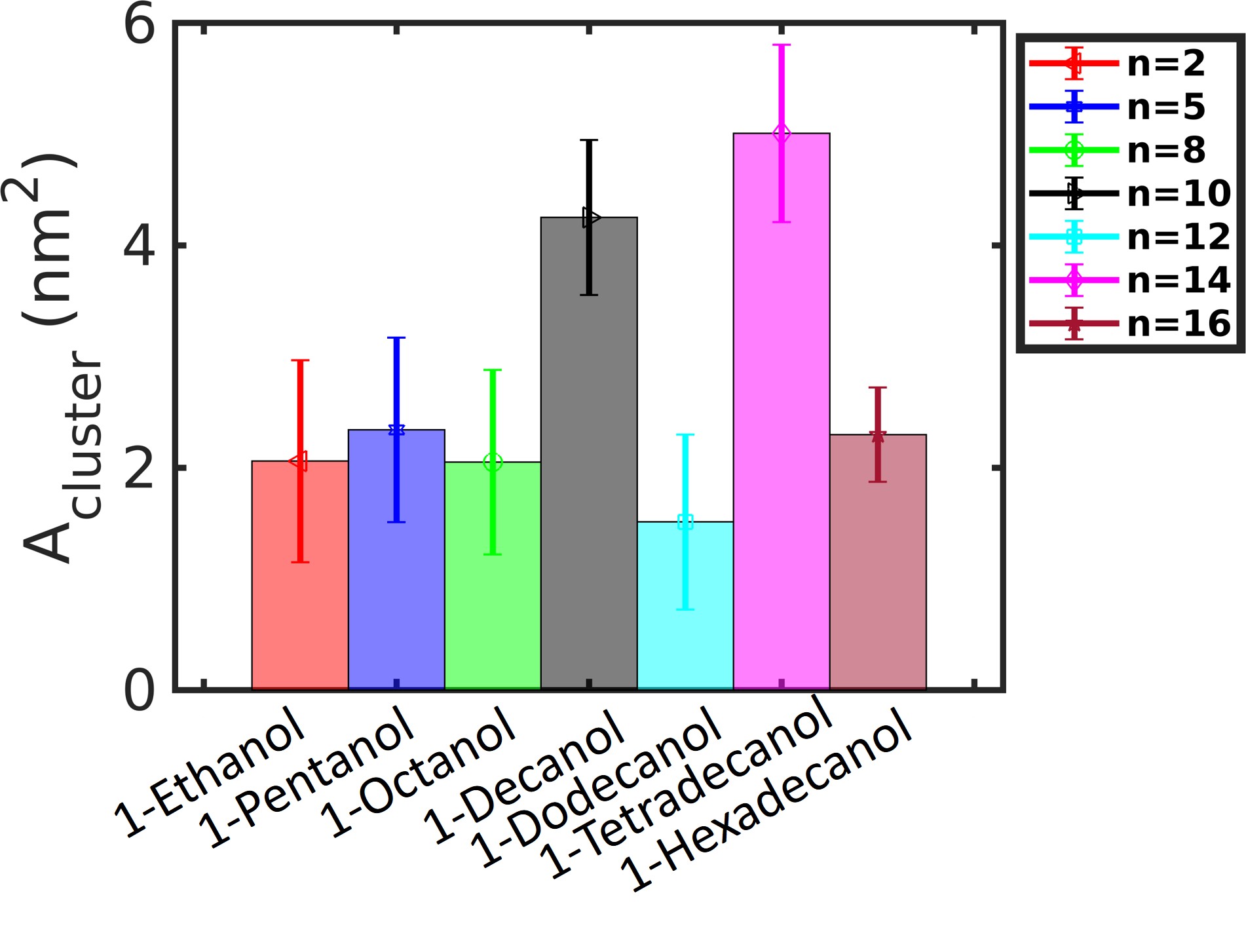}
\caption{Area of clustering, $A_{cluster}$, of 1-alkanols in the composite bilayer membrane, quantified using Voronoi tessellation–based cluster analysis. In this approach, clusters are defined as contiguous regions formed by Voronoi cells whose local number density exceeds the mean alkanol density, and $A_{cluster}$ is calculated as the total area of connected high-density Voronoi regions. The results demonstrate domain formation of 1-alkanols in all membranes, with short-chain alkanols preferentially clustering in the DOPC-rich liquid-disordered ($_d$) domains and long-chain alkanols aggregating in the DPPC-rich liquid-ordered ($l_o$) domains. Therefore, the growth and relocation of $A_{cluster}$ with chain length correlate with alkanol-induced reductions in lateral pressure and membrane elasticity, supporting a membrane-mediated origin of the anesthetic cutoff phenomenon.
}
\label{fig5}
\end{center}
\end{figure*}

To quantify the clustering behavior of 1-alkanols within the lipid membrane, we employ a Voronoi-tessellation–based analysis combined with image-processing tools implemented in MATLAB. This approach avoids the limitations of conventional grid-based spatial number density methods, in which the choice of bin size can strongly influence the inferred clustering behavior and may artificially fragment individual molecules across bins.

In the Voronoi framework, each molecule is uniquely associated with a single polygonal cell, eliminating bin-size artifacts. The method is broadly applicable to molecular dynamics and Monte Carlo simulations \cite{anirban_jpcb12,anirban_cell15,anirban_jcb14,anirban_prl16}, as well as to experimental measurements of spatial heterogeneity, such as fluorescence-based techniques \cite{anirban_cell15}.

For each system, clustering analysis is performed using the last $500$ ns of the simulation trajectory, saved at $100$ ps intervals, yielding $5000$ frames per simulation. With four independent simulations for each system, a total of $20000$ frames are analyzed. For each frame, the in-plane ($x,y$) center-of-mass coordinates of all 1-alkanol molecules are extracted.

For a given frame, a Voronoi tessellation is constructed from the ($x,y$) positions of the alkanol centers of mass, assigning a unique cell of area $A_i$ to each molecule $i$. The local two-dimensional number density associated with molecule $i$ is defined as

\begin{equation}
\rho_i=\frac{1}{A_i}
\label{gamma}
\end{equation}

The mean number density of the system is given by

\begin{equation}
<\rho_i>=\langle \frac{1}{A_i} \rangle
\label{gamma}
\end{equation}

where $<A>$ is the average Voronoi cell area over all molecules in the frame.

Voronoi cells with $A_i < \langle A \rangle$ (equivalently, $\rho_i > \langle \rho \rangle$) are classified as high-density cells, while those with $A_i > \langle A \rangle$ are classified as low-density cells. Clusters are defined as spatially connected sets of high-density Voronoi cells.

To quantify cluster size, the high-density regions are converted into binary images, where connected high-density cells are represented as white pixels and low-density regions as black pixels. The area of a given cluster $k$ is then defined as

\begin{equation}
A_{{cluster}_k}=\sum_{j \in k} A_j
\label{gamma}
\end{equation}

where the sum runs over all Voronoi cells $j$ belonging to cluster $k$. Cluster areas are computed using the bwlabel function in MATLAB, which identifies connected components in the binary images based on pixel connectivity.

This analysis is performed for all frames and independent simulations. The resulting distributions of cluster sizes are used to quantify the clustering behavior of 1-alkanols as a function of chain length, as shown in Fig. \ref{fig4}.

The probability distribution of cluster areas, $P(A_{cluster})$, is computed, and the mean cluster area of 1-alkanols is presented in Fig. \ref{fig5} showing the clusters are exhibited in all composite bilayer membrane having 1-alkanols with varying $n$.

\subsection{Lateral heterogeneity of the components in membrane}

\begin{figure*}[h!t]
\begin{center}
\includegraphics[width=14.0cm]{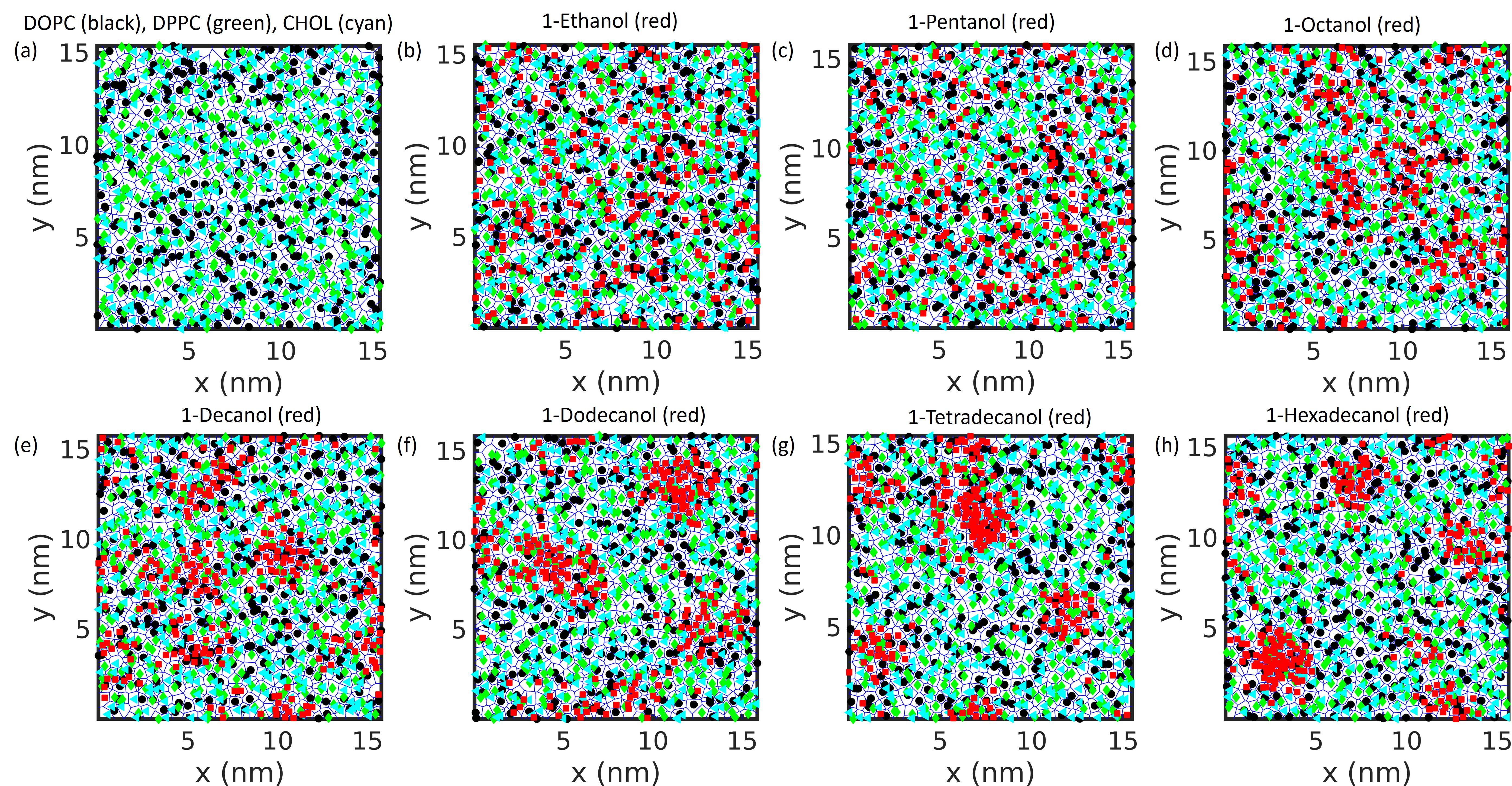}
\caption{Voronoi diagrams constructed from the centers of mass of membrane components: DOPC (black), DPPC (cyan), and cholesterol (green) for the composite bilayer (a) in the absence of alkanol, and (b–h) in the presence of 1-alkanols (red) with increasing chain length ($n=2$, $5$, $8$, $10$, $12$, $14$ and $16$), shown at $t=2000$\,ns. To examine the partitioning of 1-alkanols within the heterogeneous lipid environment, Voronoi tessellation is performed using the projected xy-coordinates of lipids and alkanols collected over the final $5000$ frames (corresponding to the last $500$\,ns) from all four independent simulation sets. From the resulting Voronoi cells, the areas associated with DOPC, DPPC, and 1-alkanols are calculated to quantify changes in local packing and lipid ordering induced by alkanol chain length. The analysis reveals that short-chain alkanols preferentially cluster within the DOPC-rich liquid-disordered ($l_d$) domains, whereas long-chain alkanols increasingly partition into and aggregate within the DPPC-rich liquid-ordered ($l_o$) domains, consistent with chain-length–dependent membrane packing and elastic responses.
}
\label{fig6}
\end{center}
\end{figure*}

To quantify the lateral heterogeneity of individual membrane components in the presence and absence of 1-alkanols, we analyze the projected center-of-mass (COM) positions of each molecular species in the membrane plane ($xy$-plane) for every time frame along the simulation trajectories. Voronoi tessellation is then constructed independently for each component, such that each Voronoi polygon contains exactly one molecule. Representative Voronoi diagrams for DOPC, DPPC, cholesterol (Chol), and 1-alkanols are shown in Fig. \ref{fig6}.

\begin{figure*}[h!t]
\begin{center}
\includegraphics[width=14.0cm]{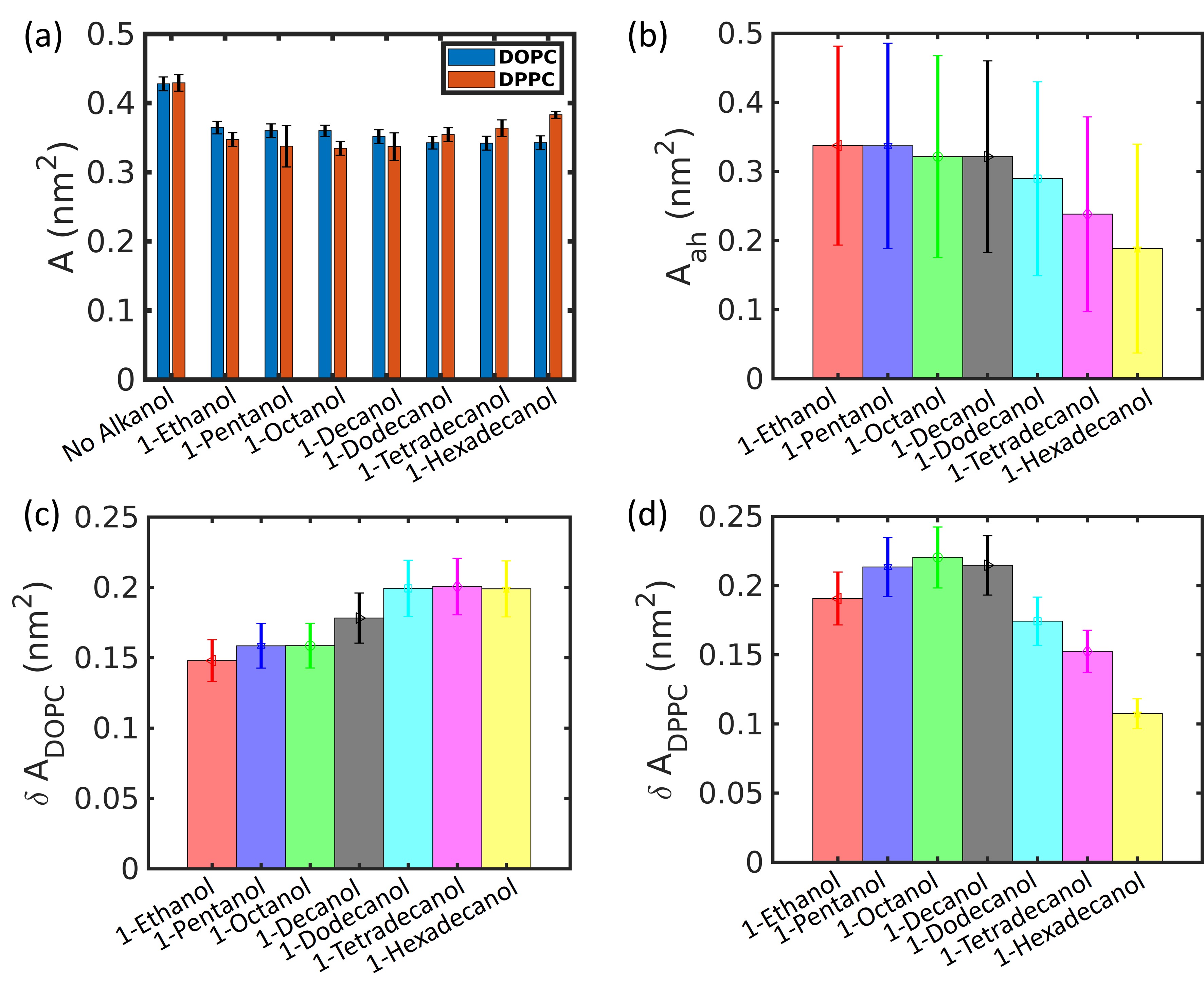}
\caption{Average molecular area of DOPC and DPPC lipids (a) and 1-alkanols (b), calculated from Voronoi diagrams constructed using the projected xy-coordinates of the centers of mass of membrane components collected over the final $5000$ frames (corresponding to the last $500$\,ns) from four independent simulation sets. Panels (c) and (d) show the relative change in the molecular area of DOPC and DPPC lipids, respectively, in the presence of 1-alkanols with increasing acyl chain length ($n=2$, $5$, $8$, $10$, $12$, $14$ and $16$). The relative area change is defined as $\delta A_{DOPC/DPPC}=\frac{A_{DOPC/DPPC}^{no\,alkanol}-A_{DOPC/DPPC}^{alkanol}}{A_{DOPC/DPPC}^{no\,alkanol}}$, where $A_{DOPC/DPPC}^{no\,alkanol}$ and $A_{DOPC/DPPC}^{alkanol}$ denote the average molecular areas of DOPC or DPPC in the absence and presence of 1-alkanols, respectively. Panel (a) shows that the average area of both DOPC and DPPC decreases upon incorporation of 1-alkanols, while panel (b) reveals a monotonic decrease in the molecular area of 1-alkanols with increasing chain length. Panel (c) shows that $\delta A_{DOPC}$ increases with alkanol chain length up to $n=12$ and subsequently saturates, reflecting the preferential accumulation of short-chain alkanols in DOPC-rich liquid-disordered ($l_d$) domains, where they most effectively reduce local lipid area. In contrast, panel (d) exhibits a nonmonotonic dependence of $\delta A_{DPPC}$, increasing for short-chain alkanols ($n=2$, $5$, $8$) but decreasing for longer-chain alkanols ($n \ge 12$). This trend indicates that although alkanols generally reduce lipid area via interfacial packing, long-chain alkanols preferentially partition into DPPC-rich liquid-ordered ($l_o$) domains and induce tail stretching and enhanced ordering, partially offsetting the area reduction. Overall, these results demonstrate a domain-specific coupling between alkanol chain length, lipid packing, and membrane elastic response.
}
\label{fig7}
\end{center}
\end{figure*}

\begin{figure*}[h!t]
\begin{center}
\includegraphics[width=12.0cm]{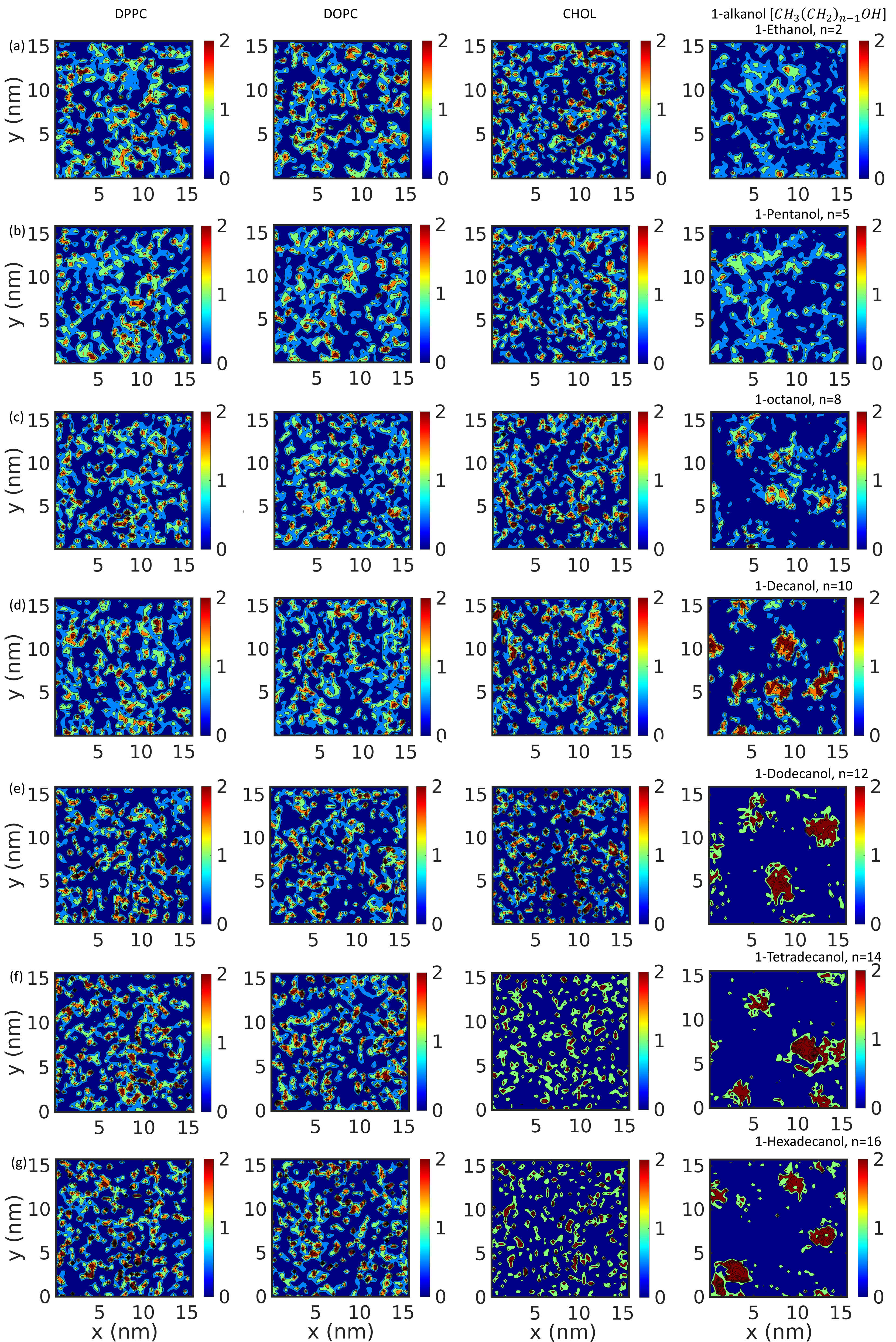}
\caption{Spatial number density maps of DPPC, DOPC, cholesterol (Chol), and 1-alkanols shown in the first, second, third, and fourth columns, respectively, averaged over the final 500 ns of the $2000$\,ns trajectories from all four independent simulation sets of the composite bilayer membrane containing 1-alkanols with varying acyl chain lengths. Number densities (in units of molecules $nm^{-2}$) are calculated using an equal square-grid approach. To identify clustering of 1-alkanols, the membrane plane ($\sim 15 \times 15 \, nm^2$) is partitioned into $15 \times 15$ square bins with a bin size of $1\,nm$, and high-density bins are identified from the corresponding look-up table (LUT) color scale in Fig. \ref{fig8}. Bins with local number density exceeding the system-wide average are classified as 1-alkanol cluster domains. All membranes exhibit coexistence of DPPC-rich liquid-ordered ($l_o$) domains and DOPC-rich liquid-disordered ($l_d$) domains. Consistent with the alkanol chain-length–dependent partitioning, short-chain alkanols preferentially accumulate within DOPC-rich $l_d$ domains, whereas long-chain alkanols selectively partition into DPPC-rich $l_o$ domains, reinforcing domain-specific packing and elastic heterogeneity of the membrane.  
}
\label{fig8}
\end{center}
\end{figure*}

\begin{figure*}[h!t]
\begin{center}
\includegraphics[width=18.0cm]{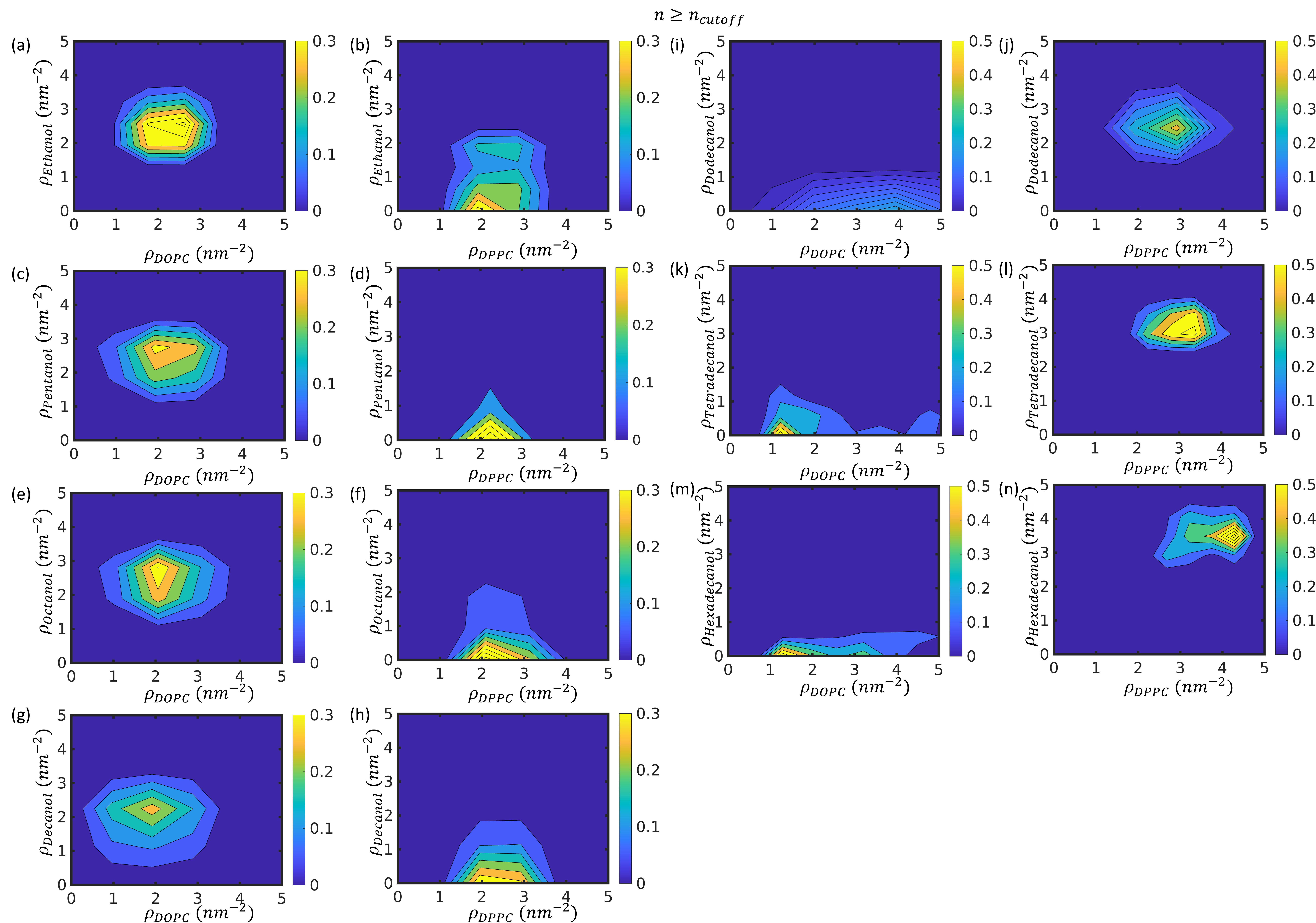}
\caption{Joint probability distributions (JPDs) of the local number density of 1-alkanols, $\rho_{alkanol}$, with those of DOPC, $\rho_{DOPC}$ (first and third columns), and DPPC, $\rho_{DPPC}$ (second and fourth columns), respectively. For short-chain 1-alkanols with chain length $n < n_{cutoff}=12$, the $\rho_{alkanol}$-$\rho_{DOPC}$ JPD exhibits a pronounced diagonal ridge, indicating strong spatial colocalization and preferential partitioning of alkanols into DOPC-rich liquid-disordered ($l_d$) domains, while no corresponding diagonal feature is observed in the $\rho_{alkanol}$-$\rho_{DPPC}$ JPD. In contrast, for long-chain 1-alkanols with $n \ge n_{cutoff}$, the diagonal ridge vanishes in the $\rho_{alkanol}$-$\rho_{DOPC}$ JPD, whereas a distinct diagonal peak emerges in the $\rho_{alkanol}$-$\rho_{DPPC}$ JPD, reflecting preferential association and clustering of long-chain alkanols within DPPC-rich liquid-ordered ($l_o$) domains. This chain-length–dependent crossover in colocalization behavior provides a direct spatial signature of the anesthetic cutoff phenomenon in heterogeneous membranes.
}
\label{fig9}
\end{center}
\end{figure*}

\begin{figure*}[h!t]
\begin{center}
\includegraphics[width=12.0cm]{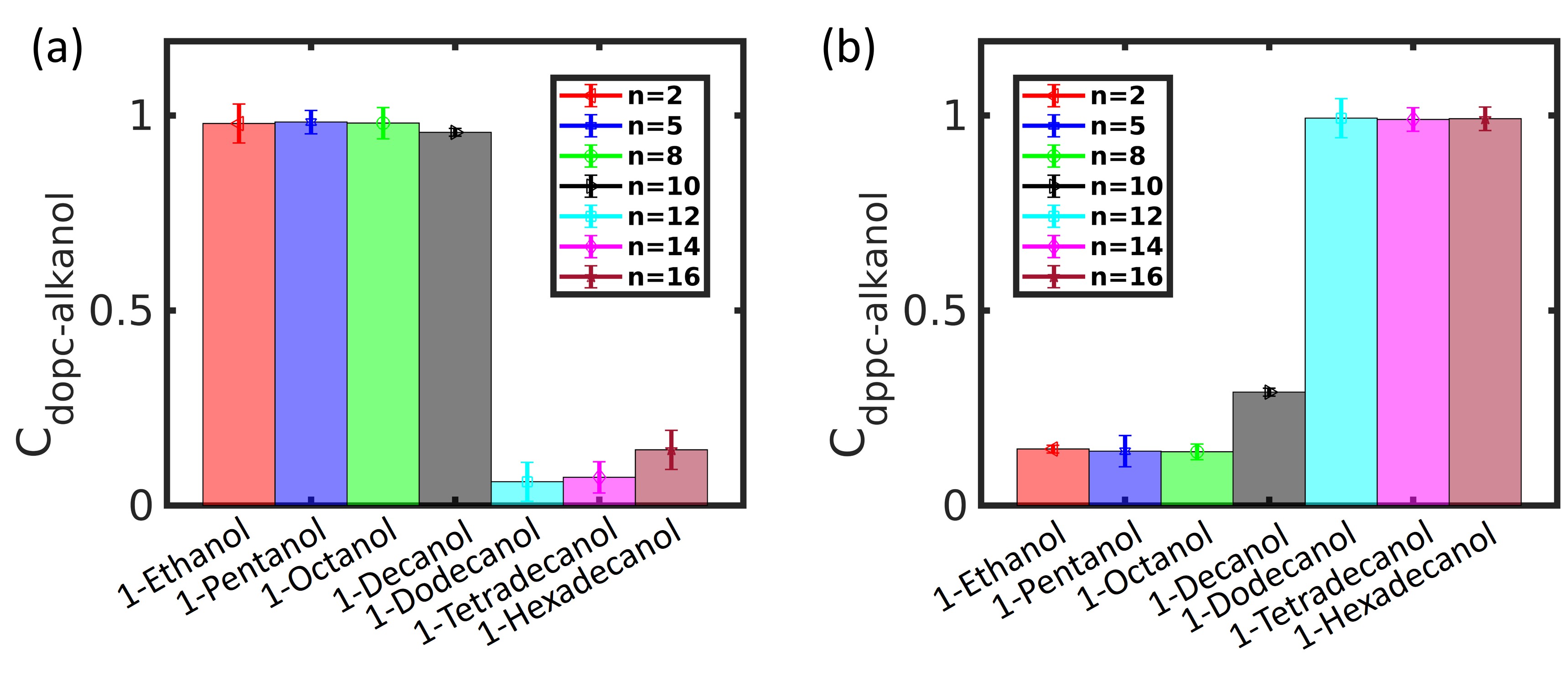}
\caption{Normalized cross-correlation coefficients between the local number densities of DOPC and 1-alkanols, $C_{dopc-alkanol}$ (a), and between DPPC and 1-alkanols, $C_{dppc-alkanol}$ (b), for composite bilayers containing 1-alkanols with increasing acyl chain length $n$. Panel (a) shows high values of $C_{dopc-alkanol}$ for short-chain 1-alkanols ($n<n_{cutoff}$), indicating strong spatial association with DOPC-rich liquid-disordered ($l_d$) domains, followed by a sharp decrease for $n> n_{cutoff}$. In contrast, panel (b) exhibits low values of $C_{dppc-alkanol}$ for short-chain 1-alkanols and a pronounced increase for long-chain 1-alkanols ($n\ge n_{cutoff}$), confirming their preferential accumulation within DPPC-rich liquid-ordered ($l_o$) domains. These cross-correlation trends quantitatively corroborate the chain-length–dependent partitioning and domain-selective clustering of 1-alkanols observed from Voronoi-based cluster analysis and joint probability distributions.
}
\label{fig10}
\end{center}
\end{figure*}

In addition to the Voronoi analysis, we compute the spatial number density of each membrane component using a conventional grid-based method. For this purpose, we analyze the last $500$\,ns of each of the four independent simulations per system. The COM positions of each component are projected onto the $xy$-plane, and the spatial number density is calculated by binning the data with a bin size of $0.1$\,nm.

From Fig. \ref{fig6}, we determine the average molecular areas of DOPC and DPPC lipids in the raft-like membrane both in the absence and presence of 1-alkanols. The results indicate an overall reduction in lipid area upon incorporation of 1-alkanols, as summarized in Fig. \ref{fig7} (a). In addition, the average area per 1-alkanol molecule decreases monotonically with increasing acyl chain length $n$, as shown in Fig. \ref{fig7} (b).

To quantify the effect of alkanol chain length on the lipid molecular area, we define the relative change in area as

\begin{equation}
\delta A_{DOPC/DPPC}=\frac{A_{DOPC/DPPC}^{no\,alkanol}-A_{DOPC/DPPC}^{alkanol}}{A_{DOPC/DPPC}^{no\,alkanol}}
\end{equation}

where $A_{DOPC/DPPC}^{no\,alkanol}$ and $A_{DOPC/DPPC}^{alkanol}$ denote the average molecular areas of DOPC or DPPC in the absence and presence of 1-alkanols, respectively.

Fig. \ref{fig7} (c) shows that $\delta A_{DOPC}$ increases with alkanol chain length up to $n=12$ and subsequently saturates. This behavior reflects the preferential accumulation of short-chain 1-alkanols in DOPC-rich liquid-disordered ($l_d$) domains, where they more effectively reduce the local lipid area. In contrast, Fig. \ref{fig7} (d) shows that $\delta A_{DPPC}$ increases for short-chain alkanols ($n=2$, $5$, $8$) but decreases for longer-chain alkanols ($n \ge 12$). This nonmonotonic trend suggests that while all alkanols reduce lipid area through interfacial packing, long-chain alkanols—preferentially partitioning into DPPC-rich liquid-ordered ($l_o$) domains—induce stretching and ordering of saturated lipid tails, partially compensating the area reduction. These results highlight the domain-dependent coupling between alkanol chain length and lipid packing in heterogeneous membranes.

The resulting spatial number density maps of DOPC, DPPC, cholesterol, and 1-alkanols are shown in Fig. \ref{fig8} (a–d). All symmetric bilayers exhibit clear phase coexistence, characterized by DPPC-rich $l_o$ domains and DOPC-rich $l_d$ domains, with cholesterol preferentially partitioning into the DPPC-rich regions where the heterogeneity of the thickness in the membrane is exhibited as shown in Fig. S6 in the SI. Notably, short-chain 1-alkanols (1-Ethanol, 1-Pentanol, 1-Octanol, and 1-Decanol) preferentially localize within DOPC-rich $l_d$ domains, whereas longer-chain 1-alkanols (1-Dodecanol, 1-Tetradecanol, and 1-Hexadecanol) exhibit preferential partitioning into DPPC-rich $l_o$ domains.

To further quantify this chain-length-dependent partitioning behavior, we compute the joint probability distributions (JPDs) of the local number density of 1-alkanols, $\rho_{alkanol}$ with those of DOPC, $\rho_{DOPC}$ and DPPC, $\rho_{DPPC}$, respectively, as shown in Fig. \ref{fig9}. For 1-alkanols with chain length $n<n_{cutoff}=12$, the JPD of $\rho_{alkanol}$ and $\rho_{DOPC}$  exhibits a pronounced diagonal ridge, indicating strong spatial colocalization, while no such feature is observed in the corresponding JPD with $\rho_{DPPC}$. In contrast, for $n\ge n_{cutoff}$, the diagonal feature disappears in the $\rho_{alkanol}$–$\rho_{DOPC}$ JPD, whereas a distinct diagonal peak emerges in the $\rho_{alkanol}$–$\rho_{DPPC}$ JPD, reflecting preferential association with DPPC-rich domains.

To provide a quantitative measure of this spatial partitioning, we define a correlation coefficient based on the normalized cross-correlation between the local number density fields of 1-alkanols and DOPC or DPPC \cite{anirban_jcb14,anirban_cell15}:

\begin{widetext}
\begin{equation} 
C(\rho^{DOPC/DPPC}(r),\rho^{alkanol}(r))= 
\frac{\langle \rho^{DOPC/DPPC}(r) \rho^{alkanol}(r) \rangle - \langle \rho^{DOPC/DPPC}(r) \rangle \langle \rho^{alkanol}(r) \rangle}{\sqrt{\langle \rho^{DOPC/DPPC}(r)^2 \rangle - \langle \rho^{alkanol}(r) \rangle^2} \sqrt{\langle \rho^{DOPC/DPPC}(r)^2 \rangle - \langle \rho^{alkanol}(r) \rangle^2}} 
\end{equation}
\end{widetext}

Here, $r=(x,y)$ denotes the in-plane position, and angular brackets, $\langle ... \rangle$ represent spatial averaging. The resulting spatially averaged correlation coefficients are denoted as $C_{dopc-alkanol}$ and $C_{dppc-alkanol}$.

The dependence of these correlation coefficients on the alkanol chain length $n$ is shown in Fig. \ref{fig10}. Fig. \ref{fig10} (a) shows a high value of $C_{dppc-alkanol}$ for $n<n_{cutoff}$, indicating strong association of short-chain 1-alkanols with DOPC-rich domains, followed by a marked decrease for $n>n_{cutoff}$. Conversely, Fig. \ref{fig10} (b) reveals low values of $C_{dppc-alkanol}$ for short-chain 1-alkanols and a pronounced increase for $n\ge n_{cutoff}$, confirming the preferential accumulation of long-chain 1-alkanols within DPPC-rich $l_o$ domains.

\section{Conclusion}


 The present results provide a unified membrane-based mechanism for the anesthetic cutoff phenomenon by linking chain-length–dependent lateral partitioning of 1-alkanols to concomitant changes in membrane mechanics, including lateral pressure profiles, compressibility, and bending rigidity. In homologous series of n-alkanols, anesthetic potency is known to increase monotonically from ethanol through dodecanol and then abruptly vanish for longer chains, despite continued increases in hydrophobicity and membrane affinity. Our simulations reproduce the molecular signatures underlying this nonmonotonic behavior.

For short-chain 1-alkanols ($n<n_{cutoff}\approx 12$), including ethanol, pentanol, and octanol, the spatial density maps, joint probability distributions, and positive correlation coefficients $C_{dopc-alkanol}$   demonstrate a strong preference for DOPC-rich liquid-disordered ($l_d$) domains. These domains exhibit reduced lipid packing, enhanced membrane compressibility, and softer elastic response. Consistent with this, the presence of short-chain alkanols leads to a pronounced reduction in the magnitude of the lateral pressure profile and its first and second moments, accompanied by a systematic decrease in bending rigidity $\kappa$ and Gaussian curvature modulus $\kappa_G$. Such mechanical softening enhances membrane susceptibility to deformation and is expected to modulate the conformational energetics of membrane-embedded proteins, in agreement with experimental observations that anesthetic efficacy correlates with membrane fluidization and pressure-profile redistribution.

In contrast, long-chain 1-alkanols ($n\ge n_{cutoff}$), such as dodecanol and longer homologs, exhibit a qualitative shift in lateral organization. The joint probability distributions reveal the disappearance of correlated density fluctuations with DOPC and the emergence of strong correlations with DPPC-rich liquid-ordered ($l_o$) domains, as confirmed by elevated values of $C_{dppc-alkanol}$. These ordered domains are characterized by high lipid packing density, cholesterol enrichment, and increased bending rigidity. Correspondingly, the lateral pressure profiles in membranes containing long-chain alkanols show substantially weaker perturbations, with minimal changes in surface tension and elastic moduli relative to the alkanol-free membrane.

This redistribution has direct consequences for anesthetic function. Although long-chain alkanols penetrate deeper into the bilayer and remain membrane-bound, their preferential sequestration within rigid $l_o$ domains suppresses their ability to alter membrane elasticity and lateral stress fields. The enhanced clustering of long-chain alkanols within these domains further reduces their effective interaction area with disordered regions, where modulation of pressure profiles and protein–lipid coupling is most efficient. As a result, beyond the cutoff chain length, additional hydrophobicity no longer translates into functional membrane perturbations.

Our findings therefore provide a mechanistic reinterpretation of the classical Meyer–Overton correlation \cite{matsumoto_bba2024}. While membrane solubility governs anesthetic potency for short-chain alkanols, the anesthetic cutoff emerges from a chain-length–induced transition in lateral membrane organization and mechanics. The cutoff corresponds to the point at which alkanols preferentially partition into mechanically rigid domains that are ineffective at mediating anesthetic action.

Taken together, these results establish that the anesthetic cutoff is not a failure of membrane binding but a consequence of domain-selective partitioning and suppressed mechanical response. By explicitly connecting alkanol chain length, lateral heterogeneity, pressure-profile modulation, and membrane elasticity, this study highlights the central role of membrane physical chemistry in anesthetic mechanisms and provides a predictive framework for understanding cutoff phenomena in complex lipid environments.

Finally, it is instructive to contrast the membrane-mediated mechanism emerging from this study with protein-centric models proposed to explain the anesthetic cutoff. Protein-based explanations typically invoke the finite size of hydrophobic binding pockets in ion channels or receptors, suggesting that longer-chain anesthetics fail to bind productively once steric constraints are exceeded. While such models may account for cutoff behavior in specific protein systems, they do not readily explain the universality of the cutoff across chemically diverse anesthetics nor its strong dependence on membrane composition and phase state. In contrast, the present results demonstrate that cutoff behavior arises naturally from collective membrane properties: beyond a critical chain length, 1-alkanols undergo a domain-selective redistribution into mechanically rigid, cholesterol-rich $l_o$ regions, where their capacity to perturb lateral pressure profiles, membrane compressibility, and bending elasticity is strongly attenuated. Importantly, long-chain alkanols remain membrane-associated and deeply inserted, yet become functionally inactive due to suppressed coupling to membrane mechanics. This distinction reconciles the continued membrane solubility of long-chain alkanols with their loss of anesthetic potency and underscores that the cutoff is governed not by molecular exclusion from the membrane or proteins, but by a qualitative shift in how anesthetics modulate membrane physical properties. These findings support a unifying membrane-mediated framework in which anesthetic efficacy is controlled by domain-specific mechanical response rather than by direct steric limitations of protein binding sites.
  
\section{Acknowledgement}
A.P. appreciates the hospitality of generous computing facilities at  SASTRA University, Thanjavur, Tamilnadu. 
A.P. acknowledges the support under Science and Engineering Research Board (SERB), Department of Science and Technology, Government of India [SERB-SRG/2022/001489] and T.R. Rajagopalan research fund, SASTRA University, India.




\bibliographystyle{apsrev4-2}



%

\end{document}